\documentclass{vgtc}                          % final (conference style)
\graphicspath{{figures/}{pictures/}{images/}{./}} % where to search for the images

\usepackage{times}                     % we use Times as the main font
\usepackage{tabu}                      % only used for the table example
\usepackage{booktabs}                  % only used for the table example
\usepackage{lipsum}                    % used to generate placeholder text
\usepackage{mwe}                       % used to generate placeholder figures

\usepackage{mathptmx}                  % use matching math font

\onlineid{4027}

\vgtccategory{Application}

\vgtcinsertpkg

\title{Audience Amplified: \linebreak
Virtual Audiences in Asynchronously Performed AR Theater\thanks{This is a preprint version of this article. The final version of this paper can be found in the Proceedings of IEEE ISMAR 2024. For citation, please refer to the published version. This work was initially made available on the author's personal website [yujnkm.com] in September 2024, and was subsequently uploaded to arXiv for broader accessibility.}}

\author{You-Jin Kim\\ %
    \parbox{1.4in}{\scriptsize \centering University of California \\ Santa Barbara \vspace{0.1cm} \\Texas A\&M University} %
\and Misha Sra\\ %
    \parbox{1.4in}{\scriptsize \centering University of California \\ Santa Barbara} %
\and Tobias Höllerer\\ %
    \parbox{1.4in}{\scriptsize \centering University of California \\ Santa Barbara}}

\teaser{
  \centering
  \includegraphics[width=\linewidth]{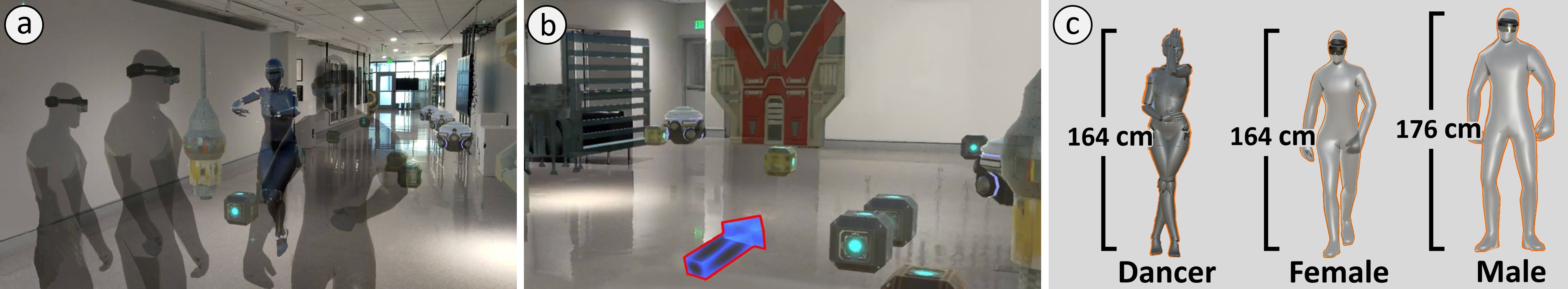}
  \caption{(a) The perspective of an augmented reality headset user witnessing a dance performance alongside virtual audience members in a physical layout with augmented stage props. (b) In every trial, a navigational aid (blue arrow) was present and guided users to the next Content Zone. The arrow vanished at the beginning of the performance and reappeared at the end of the dance. If user got lost, they could look down to find the arrow, which would lead them to their next content zone. (c) Avatars in AR Theater.}
  \label{fig:teaser}
}

\abstract{
Audience reactions can considerably enhance live experiences; conversely, in anytime/anywhere augmented reality (AR) experiences, large crowds of people might not always be available to congregate. To get closer to simulating live events with large audiences, we created a mobile AR experience where users can wander around naturally and engage in AR theater with virtual audiences trained from real audiences using imitation learning. This allows us to carefully capture the essence of human imperfections and behavior in artificial intelligence (AI) audiences. The result is a novel mobile AR experience in which solitary AR users experience an augmented performance in a physical space with a virtual audience. Virtual dancers emerge from the surroundings, accompanied by a digitally simulated audience, to provide a community experience akin to immersive theater. In a pilot study, simulated human avatars were vastly preferred over just audience audio commentary. We subsequently engaged 20 participants as attendees of an AR dance performance, comparing a no-audience condition with a simulated audience of six onlookers. Through questionnaires and experience reports, we investigated user reactions and behavior. Our results demonstrate that the presence of virtual audience members caused attendees to perceive the performance as a social experience with increased interest and involvement in the event. On the other hand, for some attendees, the dance performances without the virtual audience evoked a stronger positive sentiment.
} % end of abstract


\begin{document}


\firstsection{Introduction}

\maketitle
Mixed reality (MR) has gained significant traction due to its ability to offer a flexible and convenient way to engage with content and interact with others~\cite{feiner1997touring, yang2019dreamwalker}. Social VR gatherings and concerts (such as in VRChat) have demonstrated a desire for virtual meetups around artistic events. The COVID-19 pandemic, with its demand for isolation and personal separation, triggered the need for new creative ways to address connection. Technological advances in AR and machine learning opened pathways to experience previously recorded performances at the convenience of the participant's time and space while recreating an aura of a social gathering through AI-enhanced audience simulation. Our work demonstrates and evaluates steps in this exciting new direction of socially amplified asynchronously performed AR theater.

% talk about how it's a participatory/collective experience
In-person immersive theater has influenced the design of virtual theatrical experiences, where participants collectively view content online in a shared virtual space~\cite{game2021murder, game2021shrine}. Many of these experiences allow users to pursue various activities concurrently, all while enjoying the distributed content. This contrasts current trends in immersive theater with a directed narrative, such as productions like \emph{Dear Angelica}~\cite{game2019angelica}, which requires full immersion, undivided attention, and active interaction from the audience.

% talk about the NPCs that make it this collective experience
While having multiple users in the same space (real or virtual) is the best way to create shared participatory experiences, it could also be challenging to coordinate if users are not all participating concurrently. Including virtual characters in place of other users offers a possible solution. The usage of avatars to represent remote users or Non-Player Characters (NPCs) in mixed reality has been a prominent topic of research in recent years~\cite{piumsomboon2018minime, ho2022perspective}. Other research explored how users interact with crowds of people, walking together to understand the effects of virtual crowds on natural locomotion~\cite{trivedi2023human}. Additionally, researchers have sought to enhance the authenticity and engagement of mixed reality narratives by incorporating AI-trained characters~\cite{merrick2007reinforcement, merrick2006motivated, campo2023assessment}. Recent virtual reality games like \emph{Modbox}, allow users to interact with GPT-3 powered NPCs~\cite{article2021npc}. While AI research in Human-Computer Interaction (HCI) has seen increased efforts, its practical application in generating real-time adaptive narratives, automating character behaviors, and dynamically adjusting virtual environments based on user interactions remains relatively unexplored. Height is another intriguing factor, given that the experience of attending live concerts can be strongly impacted by the height of viewers amidst the crowd. While height manipulation in virtual reality (VR) can change stress levels in task-oriented situations~\cite{macey2023feeling}, relatively little research has been done on the relationship between user height and interacting avatar heights. As height can directly influence task-based experiences, we felt it was important to account for how this affects the desire to watch an experience.

% talk about us bringing that into MR theater
Our work adopts a participatory open-world exploration format to grant users the freedom to pursue the narrative at their own pace. This aligns with the shared VR experiences in widely used applications such as VR Chat and Rec Room. 
%We believe more and more shared experience will be like this. 
We employed AI audiences trained on human audience data to simulate an immersive AR theater/concert experience where viewers can stroll and watch dance performances with AI crowds. 
%this magical moment to always-ready experience %%made possible by human-trained AI audiences. Further, our approach aims to enhance the sense of immersion and realism, simulating the experience of attending live events and fostering a more open-world platform sensation. In this context, our work aligns with the investigation conducted by Slater et al., who explored the plausibility and illusion of realness in virtual events and situations when virtual audiences are incorporated into concert hall settings in virtual reality (VR)~\cite{mel2022sentiment}. 

While previous works investigated how crowds can be simulated and affect our experience~\cite{mel2022sentiment, yang2019dreamwalker, trivedi2023human}, we aim to bring live entertainment experiences one step closer to being ready to be experienced anytime, anywhere. We present a novel mobile AR dance performance showcase that adapts as users move through a designated area while featuring virtual dancers. This supports previous research which established that the movement and behavior of virtual audience avatars and crowds directly affect the participatory experience~\cite{yakura2020enhancing, trivedi2023human}. 

We conducted two pilot studies to determine the best appearance and behavior of the audience avatars for this experience. The findings from these studies were used to design the main AR dance performance. The main user study explored how navigation and virtual audiences affect engagement, enjoyment, and perceived realism. Users experienced the dance performance under two conditions: with and without virtual audiences. Semi-structured interviews, the SentimentAnalysis package~\cite{feuerriegel2018package}, 150-word experience essays, free-form essay responses, and analysis of questionnaire responses were employed to understand user experiences and reactions to the audience avatars. Our findings indicate that virtual AI audiences trained for specific tasks can enhance (MR) theatrical encounters, fostering a more socially engaging experience with increased participant interest and involvement.

\begin{itemize}
    \item Participants had a more positive experience when watching the performance with a virtual audience and also spent more time in this condition. This suggests increased engagement when virtual audiences are present.
    \item Participant experience differed depending on which of the two conditions they saw first, with a more positive reaction overall when virtual audiences were included after first watching the performance alone. This indicates the importance of careful introduction of characters into the interactive narrative.
    \item The user's height emerged as a factor influencing their overall experience. This finding highlights the importance of considering physical attributes and user perspective in the design of immersive MR experiences.
\end{itemize}

We see great possibilities for AR performances that can be experienced via mobile AR by individual audience members at their leisure, while still experiencing some form of audience connection. In the future, such an audience member's social viewing behavior could influence the performance experience of subsequent users.

\section{Related Work}
This section reviews immersive theater, the usage of NPC and ML-trained avatars, and user sentiment.

\subsection{AR Immersive Theater}
Recently, there has been a shift toward bringing theater experience to the mixed reality. Despite some overlap with traditional theater production, MR theater development presents novel challenges and opportunities~\cite{coulombe2021virtual, VR_SharedSocialSpaces}. Our research draws inspiration from narrative theater experiences, such as \emph{CAVE}~\cite{CAVE_Leon}, \emph{CAVRN}~\cite{herscher2019cavrn}, \emph{Gulliver}~\cite{ARGulliver}, and the integration of physical props with AR features~\cite{nicholas2021expanding}.
Our project builds upon these foundational works to create an interactive user-centered dance performance experience through walking. This aligns with our broader research focus on crafting experiences for specific locations and spaces~\cite{wang2022exploring, kim2023rdr}. Notable examples of location-based experiences include \emph{Soul Hunter}~\cite{weng2011soul}, \emph{ARQuake}~\cite{thomas2000arquake}, \emph{HoloRoyale}~\cite{rompapas2018holoroyale}, and exploration-based horror games~\cite{madsen2022fear}. The use of walking experiences in mixed reality extends to the engagement with historical events~\cite{fujihata2022behere, dow2005exploring}.

Since the research on outdoor exploration in wide-area virtual theater performances by Cheok et al. in 2002~\cite{cheok2002interactive}, various mixed reality theaters, such as \emph{Holojam in Wonderland}~\cite{gochfeld2018holojam}, \emph{The Meeting}~\cite{pietroszek2022meeting}, \emph{Dill Pickle}~\cite{pietroszek2022dill}, \emph{Samuel Beckett’s Play,}~\cite{Play_Leon} and \emph{Gumball Dreams}~\cite{lyons2023gumball}, have provided unique opportunities for active user engagement within the theater performance itself. These examples indicate the increasing importance of a player's involvement in production and development. Moreover, there is a growing trend of collective experiences where users gather to participate in interactive narratives, live concerts, and other events, sharing virtual spaces and moments~\cite{game2023japan, game2020capsule, CAVE_Leon}.

Within open-world platforms like VR Chat, users generate novel virtual spaces that provide theater-like experiences with immersive narratives~\cite{game2021murder, game2021shrine, game2023flow}. These user-generated spaces and experiences are shared through various platforms such as Reddit, YouTube, and VR infusers highlighting the growing popularity of user-generated content in the VR space~\cite{web2023reddit, web2023YT2}.

\emph{The Under Presents}~\cite{game2019presents}, a multiplayer VR game and performance space, enables users to collectively view live or recorded performances, encouraging a shared virtual experience in a participatory theater experience. Although experienced entirely on the user's own, these shared spaces where audiences build experiences together are on the rise. The deployment of AI-trained audiences in our work, intended to capture the essence of live events within our interactive AR theater experience, is said to have been directly influenced by user engagement.

\subsection{The Usage and Effect of Virtual Agents}

In mixed reality environments, NPC audiences have traditionally found their primary use in video games, with some research dedicated to understanding their influence on user experiences~\cite{10.1145/2967934.2968092}. Emmerich and Masuch extended their research~\cite{watchmeplay} to examine the impact of real observers and virtual agents on gaming experiences, while Yee et al. \cite{10.1111/j.1468-2958.2007.00299.x} conducted two studies to investigate the effect of transformed self-representation on user behaviors. Similarly, Leyrer's study \cite{10.1145/2077451.2077464} showed that avatars could significantly influence the verbal estimates of egocentric distances during the tasks and the eye height had a significant effect on the verbal estimates of both egocentric distances and the dimensions of the room. Zhu et al. \cite{zhu2023} also determined that the appearance of interactive conversational avatars impacts user experience, comfort, and the ability to recall information from the AR experience.

A recent study by Kao~\cite{kao_effects_2021} further contributed to the exploration of virtual audiences by proposing the intentional incorporation of observation and surveillance based on text phrases within games. This approach demonstrated the potential to enhance players' performance, overall gaming experience, and motivation. Notably, Haller et al. \cite{10.1145/3290688.3290752} revealed that NPC audiences who cheered and applauded led to improved player performance. In addition, Xu et al.\cite{10.3389/fpsyg.2023.1079132} conducted research to investigate the effect of the size of NPC audiences and their feedback on user performance and gameplay experience. Their results illustrated that as the size of NPC audiences grew larger, the user's performance and gameplay was enhanced. Similarly, Yu et al. \cite{Yu2023} found that the presence of NPC audiences and their feedback can enhance elderly users' performance and gameplay experience, like competence, immersion, and intuitive controls. 

While the impact of virtual audiences in some video games and MR contents has garnered increasing attention, it remains relatively understudied in virtual environments and AR theater experiences. Consequently, we would like to understand the role and effect of virtual audiences on users' performance and experience in the AR theater environment.

\subsection{Evaluation of User Experience}
Assessment is important to understanding user experiences within virtual environments and AR settings. Here, we present an evaluation of methodologies crucial to our experiment.

There are different definitions of presence from previous research \cite{10.1162/105474698565686, 10.1162/pres.1992.1.4.482}. Our study aligns most with Witmer's definition \cite{10.1162/105474698565686}, which defines presence as a subjective experience of being in one place or environment. He also stated that the user's self-reported sense of presence is an important metric to evaluate the effectiveness of virtual environments. Presence is evaluated by questionnaire \cite{lessiter2001cross, usoh2000using, 10.1162/105474698565686}, physiological responses~\cite{meehan2002physiological}, and in some cases using behavioral indicators~\cite{kisker2021behavioral}. In our study, we use a combination of post-trial questionnaires, with questions from the Game User Experience Satisfaction Scale (GUESS) \cite{doi:10.1177/0018720816669646}, the Immersive Virtual Environment Questionnaire \cite{10.1145/2927929.2927955}, the Augmented Reality Immersion (ARI) Questionnaire \cite{GEORGIOU201724}, and the Game Engagement Questionnaire (GEQ)~\cite{geq2013}, to assess user interest and boredom. Additionally, we utilize the Presence Questionnaire (PQ) \cite{10.1162/105474698565686} to collect quantitative feedback on the sense of presence experienced by participants in each trial letting them write 150 words or more about the experience and talk to interviewer freely for additional 5 minutes; gathering insights and reasoning behind the user experience (see also Section~\ref{procedure-pilot2}.). 

The concept of immersion is often employed in VR narrative games~\cite{GEORGIOU201724}. Brown et al. \cite{10.1145/985921.986048} proposed a division into three levels, including engagement, engrossment, and total immersion. Georgiou et al. \cite{GEORGIOU201724} applied the same divisions to develop and validate the Augmented Reality Immersion (ARI) questionnaire, which is one of the main sources for building our questionnaire. This notion of immersion is not to be confused with Slater's definition of immersion, which refers to objectively measurable characteristics of technology, but instead aligns more with his definition of Presence~\cite{slater2003note, slater2023sentiment}. User sentiment is another important assessment.

\section{Immersive Theater Design}
Our immersive theater concept transforms a 208.54$m^{2}$ corridor into a spaceship's hallway where users are free to walk about while watching a performance of a dancer through the Microsoft HoloLens-2 Mixed Reality headset. This section details the user experience of our study and explains how our dancer performance was recorded, along with how the ML audience was trained. 

\subsection{Location-based Immersive Theater}
We utilized the Unity game engine to develop our immersive AR theater experience featuring three content zones that showcase prerecorded contemporary dance performances through volumetric avatars. The experiment was conducted in an indoor space where users could move among the evenly distributed content zones with the help of a navigation aid. Upon reaching each zone, the corresponding recorded performance was played. Each session was designed to allow users to explore for approximately 4 minutes. 

\subsubsection{Navigational Aid}
We added a guidance system to help users locate the next content zone (dance performance) and navigate the environment at their own pace. To streamline the experience, a guiding system with an arrow pointed toward the next destination (Figure \ref{fig:teaser}b).

\subsubsection{Dancer Performance}
In a traditional theater setting, spotlights draw attention to performers. However, in an open-world performance such as ours, we needed to find new ways to draw attention to the dancer and alert viewers that the performance was in progress. 

The dancer's performances utilized motion capture data collected by Morro Motion technology~\cite{web2023mocap} using a Vicon motion capture system, consisting of six strategically positioned Vicon T160 cameras. The dancer was fitted with 36 reflective ball markers, allowing for the recording of joint trajectories and muscle activities~\cite{wojtusch2015humod}. All dancers are standardized to a height of 164$cm$. We used 34-bone avatar for the dancers in the pilot study and 102-bone avatar for the main study (Figure \ref{fig:teaser}c) as the pilot studies indicated that the dancers' movements appeared rigid. We used the same dance motion animation in both studies. Yet, the avatar with more bones could convey expressive movements, which enabled us to eliminate ``stiffness."

\subsubsection{Physical and Virtual Layout}
A digital twin of the corridor was used to handle occlusions on the headset and to train the ML audiences. To ensure precise content alignment on HoloLens 2 devices, we used Azure Spatial Anchors~\cite{buck2022azure}. These spatial anchors from the server were automatically loaded to position digital content, addressing occlusions caused by physical walls. We strategically placed three 2.8 $m^{2}$ content zones to act as location-based triggers along pathways from the user. These zones triggered dance performances when users entered specific areas on the floor layout. The zones were eight meters apart. 

\subsubsection{User Experience and Instruction} \label{instruction}
Our user experience was crafted to promote exploration, enabling the user to navigate and watch dance performances for approximately 4 minutes as they wished. After watching all three dance performances in the trial, users could continue exploring the space or exit the hallway to conclude their experience. Participants were asked not to run, and to wait for each dance performance to end before moving on. They were told about the arrows that could guide them if they were lost, and that they were free to exit the hallway when they were ready to end the experience. They were not given any details of the virtual audience(s), to prevent participant bias or preconceived notions.

\subsection{Virtual Audience}
With an understanding of the physical layout, our virtual audiences imitated the behavior of a real audience. Virtual audience agents are spatially aware, walking and observing performers with similar curiosity and attention as human users. We trained the audience agents to blend into the user's environment and experience to ensure the focus remained on the dance performers. We designed all six virtual agents to spawn close to the ticket booth, which indicated they were fellow audience members. 

\subsubsection{Training Workflow}
We employed a combination of imitation and reinforcement learning for our virtual audience. First, we collected tracking data of users navigating the AR theater environment while watching performances. This data served as the basis for imitation learning, where a teacher agent performed the task and a student agent imitated it. Our goal was to achieve human-like behavior rather than machine-like perfection. Each deep neural network (DNN) model underwent 2 million training steps using the same demonstration data. We identified the top-performing 6 models by selecting those that exceeded the threshold set at the top 30 percent of cumulative reward points.

We utilized a deep neural network with three hidden layers, each consisting of 128 neurons. The neural network was trained using the proximal policy optimization (PPO) algorithm, implemented through Unity's open-source project, Machine Learning Agents Toolkit (ML-Agents).

The virtual audience avatars were trained to move within the hallway, exploring content zones and mimicking real human behavior. We trained six neural network models using position data from previous users, which included 60 trials and 3 hours and 21 minutes of tracking data. To ensure accuracy, we used a digital twin model of the hallway at the precise scale and matched it with the demonstration data, creating a realistic training environment. These agents operated independently but shared the same model, facilitating parallel training in 18 AI environments. Using the digital twin allowed us to capture nuanced behaviors specific to our scenario, rather than relying on a generic model.

\subsubsection{Reward System}

To enhance our results from imitation learning, we fine-tuned the reward system to incentivize behavior aligned with our task of watching performances while walking. AI agents received rewards for various actions related to content zones, such as moving towards them, entering them for the first time, and remaining within them. The longer an agent stayed within a content zone, the more rewards it accrued, up to a maximum duration matching the performance duration in each zone.

To optimize the timing of engagement with the virtual audience, we introduced additional rewards. Entering the first content zone resulted in a reward, with increasing rewards for subsequent zones. Completing entry into all three zones earned the agent an extra reward. Additionally, agents received a small incentive for proximity to the content zone but were penalized for touching the wall.

Agents received positive rewards for various actions, including moving towards content zones, entering them for the first time, and staying within them. The longer they stayed in a content zone, the more rewards they earned, up to a maximum of 17 seconds, which matched the duration of the performance occur in each content zone. To refine the timing of virtual audience, we introduced additional rewards. Entering the first content zone was rewarded with 48.2f, the second with 63.7f, and the third with 85.5f. Completing entry into all three zones earned the agent an extra 41.0f. We also provided a small incentive for getting closer to the content zone (0.03f/ sec) and penalized the agent for touching the wall (0.01f/ sec).

\subsubsection{Avatar Motion}
\label{sec:logic}
We placed particular emphasis on improving avatar motion for mixed reality interactions, recognizing the significance of avatar representation in such scenarios based on existing research findings in recent years~\cite{yakura2020enhancing, doi:10.1080/10447318.2022.2121038, wang2020effect}. We leveraged Unity ML-Agents' ActionBuffers function to handle continuous and discrete actions. This approach significantly smoothed the motion during initial training sessions and prevented overlapping actions.

To provide a more realistic representation of audience movement, we took an additional step by incorporating motion-matching technology, specifically Kinematica. This technology enabled seamless transitions between different behavioral motions based on the path and speed of the avatar's movement. Our avatar animations were controlled through State Machines, encompassing gestures like walking, idle, turning around, and subtle expressive motions, thereby enhancing the overall realism of the avatar's movements.

\subsubsection{Virtual Audience Voice Responses}
We collected voice recordings from participants who had previously experienced the same virtual dance performance. To encourage users to express themselves more naturally, we assured them that speaking loudly during the performance was acceptable. To allow participants to react and express themselves freely while being voice recorded, the chatter of four planted audience members was played in the background. These recorded voice reactions were then played back as virtual audience response voices. This decision was motivated by real-world experiences and the way users of VR social experiences can always hear the people around them. 

\section{Experiments}
To optimize our study's ability to gauge the impact of virtual audiences on user experience, we conducted two pilot studies. In the first, 20 participants ranked preferred avatar and behavior combinations. In the second, based on the insights gained, 10 participants evaluated four conditions. Finally, our main study, involving 20 participants, focused on two trials: dance performance only and performance with an ML avatar as the virtual audience. The participants for the three experiments (two pilots, one main) were drawn from separate populations. No participant took part in more than one experiment. We recruited 50 participants for our experiments.

\begin{figure}
 \centering % avoid the use of \begin{center}...\end{center} and use \centering instead (more compact)
 \includegraphics[width=\columnwidth]{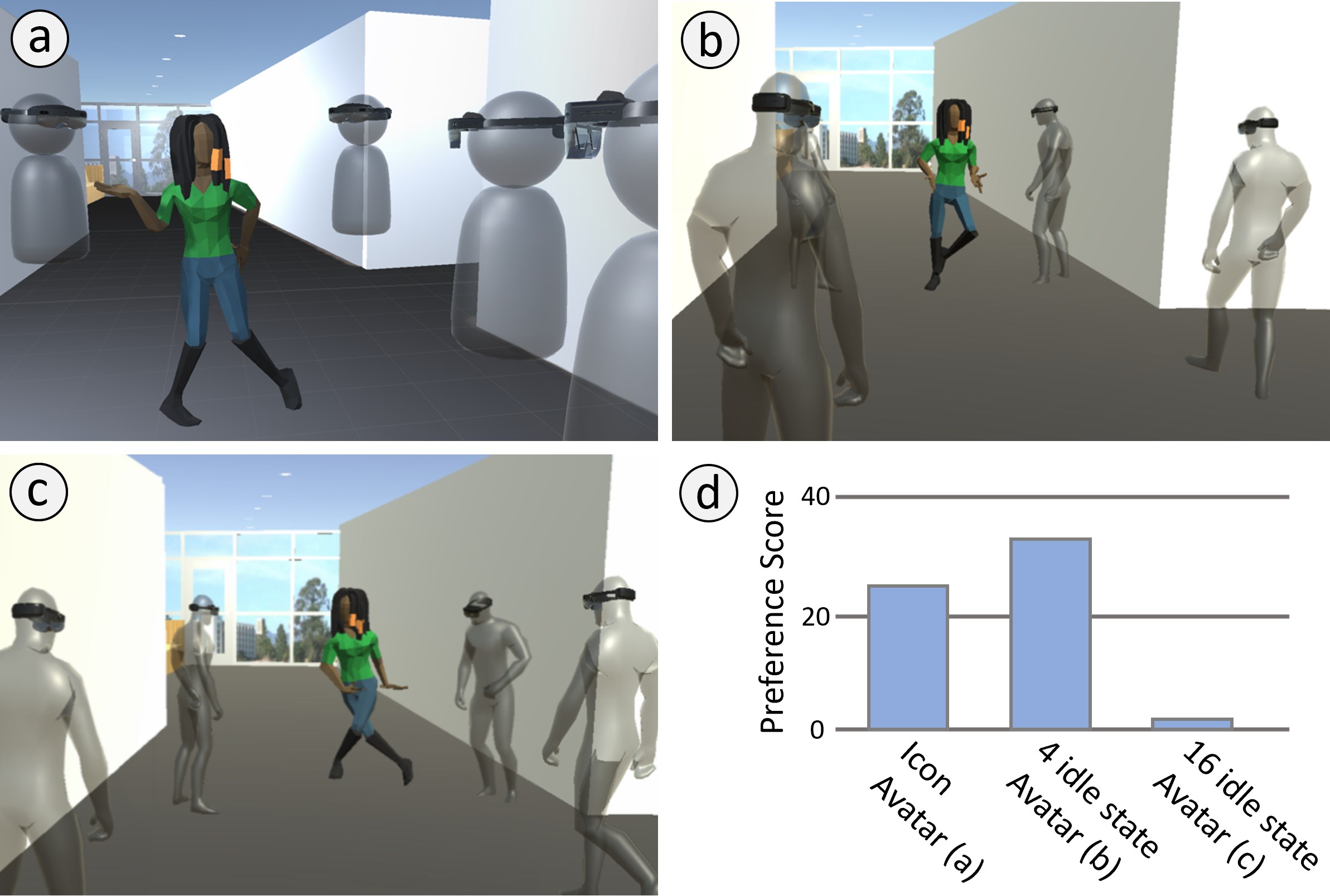}
 \caption{Users' perception of watching a dance performance with each of the three virtual audiences in the first pilot study: (a) A human figure made of basic geometric shapes represents the virtual audiences we named Icon Avatar. (b) A human figure avatar with 4 idle states represents virtual audiences. (c) A human figure avatar with 16 idle states, that enable the avatar to be more expressive and receptive, represents virtual audiences. (d) Accumulated preference ranking for each of the 3 avatars as virtual audiences (2 for the most preferred avatar, 1 for the next, and 0 for the least preferred, for a total of 20 participants). The avatar with 4 idle states was most preferred, followed by the basic geometric avatar, Icon Avatar. }
 \label{fig:pilot-study1}
\end{figure}

\begin{figure*}
\centering
  \includegraphics[width=0.8\textwidth]{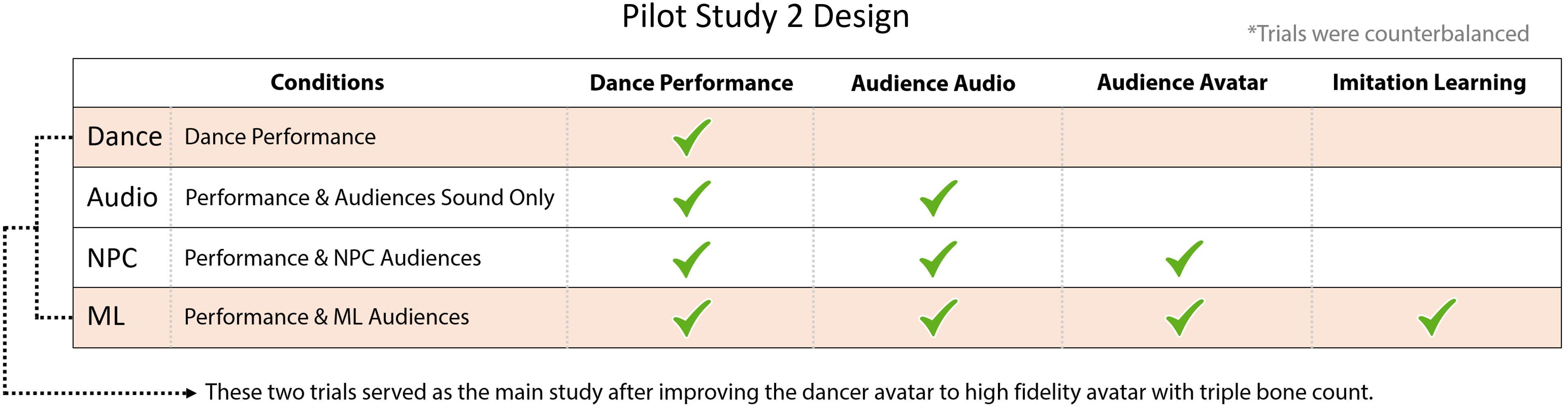}
\caption{Design chart for Pilot Study 2, showing the four trial conditions and their components. }
 \label{fig:pilot-study2}
\end{figure*}

\subsection{Pilot Study 1}
The pilot study was used to select the best avatar models to use in the second pilot study and the formal user study. For this purpose, we let the 20 participants experience different avatar types to participate in viewing the performance. Our designed AR theater environment included (1) an icon avatar, (2) a 4-idle-state avatar, and (3) a 16-idle-state avatar, which appeared more expressive than the 4-idle-state avatar (Figure \ref{fig:pilot-study1}). 

As we briefly explain in the Avatar Motion section, in the first pilot study, condition 1, our virtual audience was represented as human figures without hands or legs. As seen in Figure \ref{fig:pilot-study1}a, this was to give the viewer an idea that someone is watching nearby and nothing more. For conditions 2 and 3, we produced a human figure with arms and legs, three male avatars, 176$cm$ in height, and three female avatars of the same height as the dancer (164$cm$). For condition 2, we deployed four idle animation states on top of basic navigational motion as seen in Figure \ref{fig:pilot-study1}b. For condition 3, we deployed sixteen idle animation states on top of basic navigational motion and demonstrated much more expressive reactions, as illustrated in Figure \ref{fig:pilot-study1}c.

%%All three condition avatars appear to wear HoloLens 2 headsets to indicate they are fellow audience members. The audience avatar was shaded to 50\% transparency so that viewers may feel the audience's presence in their periphery while still being able to see the performance without occlusion should the avatar be in the way.

\subsubsection{Pilot Study 1 Procedure} 
We recruited 20 participants, aged 18 to 51, including 10 self-identified males and 10 self-identified females, from a university campus. These individuals took part in a survey involving the evaluation of a 1-minute performance under three distinct conditions within a physical environment while using an AR headset. The sequencing of the conditions was counterbalanced across participants. %%We did not include the natural locomotion component because the objective was to accurately assess the user's preferred avatar styles when they saw virtual dance performances with other virtual crowds.

Participants ranked their preferred audience from most to least desirable and participated in brief exit interviews. They discussed their ranking preferences for watching performances, the types of audiences, and which experience most resembled a live concert, explaining their favorite and least favorite choices. During the survey, participants experienced three trials in which they were presented with distinct virtual audience types while the dance performance remained consistent. Additionally, we gathered subjective feedback and suggestions from participants to enhance our future studies.

For ranking, a preference score system was employed: the first choice received 2 points, the second choice was awarded 1 point, and no points were allocated for the last choice (Figure \ref{fig:pilot-study1}d).

\subsubsection{Insight and Qualitative Feedback}
Our initial pilot study outcomes indicated that viewers exhibited a preference for the 4-idle-state avatar model as their virtual audience, followed by the icon avatar. Conversely, the expressive 16-idle-state avatar model was generally not favored, as observed in Figure \ref{fig:pilot-study1}d. The main reason cited for this preference was that fellow audience members in the expressive avatar model appeared too distracted and moved around too much, detracting from their ability to focus on the dancer's performance. In a direct quote, participant P17 expressed a preference for the 4-idle-state avatar, stating, \textit{``I am much better at maneuvering around crowds of people and need to be able to see how they are walking, so I prefer the avatar with arms and legs"}. P4 remarked, \textit{``It makes more sense for the audience to be more like the performers, as it seems odd to me that the two should not match. Additionally, a more intricate human figure simply fits in with the surroundings"}.

In light of these findings, we opted to use the 4-idle-state avatar model in our NPC and ML trials in the second pilot study.

In addition to these findings, the subjective feedback gathered from all participants after the study indicated that many participants described the 16-idle-state avatar model as ``creepy." Consequently, based on these results, we decided to utilize an audience avatar with arms and legs, the 4-idle-state (less expressive) avatar model for our main user study.

\subsection{Pilot Study 2}

\begin{figure}
 \centering
 \includegraphics[width=\columnwidth]{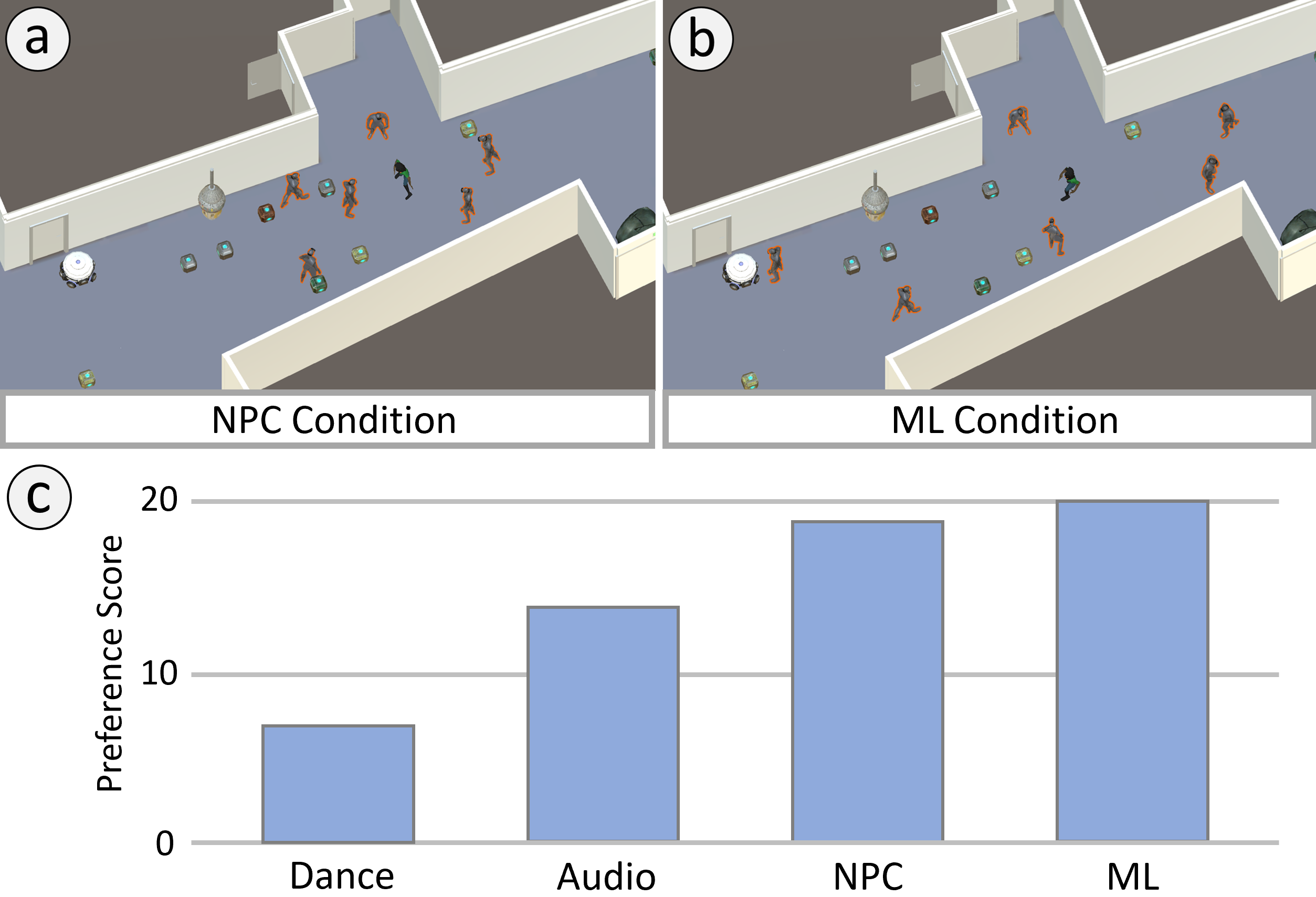}
 \caption{The positioning and dispersion of virtual audiences and the differences between the ML and NPC models. (a) Demonstrates how audiences are positioned close to the dancers and are looking in the dancer's direction. (b) Illustrates how the crowd is dispersed and spaced out, with some people watching the dancers and others moving independently. (c) Accumulated preference rankings for each of the four experiment conditions in the second pilot study (with 3 points for the most preferred and 0 points for the least preferred, from a total of 10 participants).}
 \label{fig:virtual-audiences}
\end{figure}

Pilot study 2 was conducted to determine which of three different implementations of the virtual audience to use in the main study. The control condition was the dance performance with no audience (hereafter referred to as \textit{Dance}). All trials included audio, except for the Dance condition.  The three different audience implementations are detailed below (Figure \ref{fig:pilot-study2}):

\begin{itemize}
\item\textit{Audio}: In addition to the dance performance, users also heard recordings of six real user reactions to the performance, implemented as spatial sound.

\item\textit{NPC}: In addition to the dance performance, users also saw six humanoid avatars as virtual audience members, along with the same audio reactions as \textit{Audio}. These avatars always stood close to a content zone and directly faced the performer (Figure \ref{fig:virtual-audiences}a).

\item\textit{ML}: This condition was similar to the NPC condition, except that the audience avatars were now trained to emulate human-like movement and behaviors (see Section \ref{sec:logic}). The difference between the NPC and ML audiences was that the NPC audience tended to spend their time looking directly at the performance and paying attention to it, while the ML audience typically displayed more individual behaviors, and acted more participatory and dispersed (Figure \ref{fig:virtual-audiences}b).
\end{itemize}

\subsubsection{Pilot Study 2 Procedure}\label{procedure-pilot2}
Ten participants (5 self-identified as male, and 5 as female) experienced the four conditions in a randomized order (counterbalanced between subjects) and gave feedback on their experience and preference. Each condition lasted approximately four minutes, though participants were free to explore the space for longer if they wanted to. Demographic information was collected before they experienced any of the conditions, and qualitative feedback was collected after each condition (specific to that experience) as well as after the entire study (regarding the overall experience).

The demographic questionnaire had questions related to previous experience with immersive theater, narrative games, and VR games. The post-trial questionnaire, which was administered after each condition, included a five-question sub-scale of the Game User Experience Satisfaction Scale (GUESS) \cite{doi:10.1177/0018720816669646} to measure \textit{enjoyment}, eleven questions from a sub-scale of Immersive Virtual Environment Questionnaire \cite{10.1145/2927929.2927955} to measure \textit{presence}, a twelve-question sub-scale of the Augmented Reality Immersion (ARI) Questionnaire \cite{GEORGIOU201724} to measure \textit{immersion}, and a ten-question sub-scale of the ARI Questionnaire \cite{GEORGIOU201724} to measure \textit{engagement}, three questions from Game Engagement Questionnaire (GEQ) to measure interest and boredom~\cite{geq2013} . Participants also completed a free-form short essay (150 words or more) within 10 minutes about their experience, along with four 7-point Likert scale responses mentioned above. At the end, they engaged in a 5 minute voice recorded free talk for analysis and insight. %%The post-experience questionnaire included 21 questions about their overall experience and impressions of the stage design, dance performance, and virtual audience, all measured on a 7-point Likert scale.

\subsubsection{Insight and Qualitative Feedback}
Using the SentimentAnalysis package, we generated a word cloud visualization to identify trends in user experience~\cite{ALAMOODI2021114155}. Additionally, we employed Grounded Theory for our analysis, involving validation distributed among different team members~\cite{corbin2014basics}. This process allowed us to review the user interviews, discuss our findings collaboratively, and compile an overview of the general trends, which we report here. The ML condition emerged as the most preferred condition, with the NPC condition being second-most preferred. Both conditions were significantly more entertaining and enjoyable than the Audio and Dance conditions (Figure \ref{fig:virtual-audiences}c). The ML condition also fostered better presence and engagement compared to NPC, with participants saying that \textit{``I liked the second one (ML) because I can walk behind them, feeling like I am watching with a cloud of people and not by myself" (P2)} and \textit{``The audience naturally guided me to the dancer without needing to rely on an arrow which was nice." (P5)}. Seven out of 10 participants reported feeling ``distracted" by NPC avatars, with some expressing that they felt ``crowded" and that their view was ``blocked," leading to occasional overwhelm. The ML avatars, on the other hand, helped to improve the experience by their tendency to disperse in the environment (\textit{``I like that people are there but I prefer the one (where) user(s) are more spread out, I can focus on dance performance without feeling like I have to fight for the best view while still feeling like I am in (a) concert," P6}).

\subsection{Main Study}
For the main study, we recruited 20 adult participants (N=20, 9 who self-identify as female, 11 who self-identify as male, ranging in age from 18 to 53, an average age of 26.25 (SD=8.37) ) from a local university campus. They were compensated at a rate of \$15 per hour. All participants were able-bodied without hearing and vision issues and were able to move around during the trials.

\subsubsection{Experimental Design}
Based on the results of our two pilot studies, we decided to use the 4-idle-state avatar model in the main study (as shown in Figure \ref{fig:teaser}b). We also opted to utilize machine-learning-trained virtual audiences, as we believe they closely mimicked a live audience experience. This main study followed a within-subject design with one factor--the presence of NPC audiences, which is evaluated with two conditions: \textbf{Dance} is the baseline condition where no audience was provided during trials and \textbf{ML} with six Imitation Learning-Trained NPC audience avatars presented during the AR theater play. Trials with virtual audiences included audio, whereas no audio was present in the Dance condition as there was no virtual audience in those trials. The order of these two conditions was counterbalanced in the experiment.

\subsubsection{Measurement}
Despite eliminating two trials compared to pilot study 2, we utilized the same set of questions and interview process. As for outcome assessment, we collected the following data to evaluate users' overall experience during trials:

\begin{itemize}
    \item User Performance: We collected the completion time of each trial for each user.

    \item User Experience: As in pilot study 2, we measured and recorded participants' experience via our designed questionnaire (see Section \ref{procedure-pilot2}). We focused on measuring four main metrics of user experience, including  \textit{Enjoyment}, \textit{Presence}, \textit{Immersion}, and \textit{Engagement}.

    \item User Sentiment: Users were required to complete a brief text-written essay with open-ended questions targeting their overall experience and feelings in each trial. For the user sentiment analysis, we chose to use the R package sentimentr \cite{ALAMOODI2021114155}, the VADER system, the syuzhet package, and SentimentAnalysis package \cite{feuerriegel2018package} to analyze user sentiment level of each trial based on what we collected from participants' answer.
    
\end{itemize}

Users also responded to questions regarding their subjective preferences and feedback about the two trials, with some open-ended questions from Presence Questionnaire (PQ) \cite{10.1162/105474698565686}, such as, \textit{``Which trial did you enjoy more?"}, \textit{``Which trial felt more like a live performance?"}. Semi-structured interviews were conducted after the entire experiment, and users were allowed to share any thoughts in a free-form interview post trial to understand the data. 

\subsubsection{Main Study Procedure}

The procedure of the main study mirrors the pilot study 2 (detailed description in Section \ref{procedure-pilot2}), with the only difference being that there are only two conditions in the main study. Thus, the entire procedure of the main study lasted approximately 60-80 minutes per participant. 

\subsubsection{Analysis}

The Shapiro-Wilk test was used to check for violations of normality in the data. The Aligned Rank Transform (ART) \cite{Jacob_ART} was applied to all data that violated normality. 

One-way repeated measures ANOVAs were used to assess the impact of the two experiment conditions (Dance and ML) on user experience ratings and metrics such as completion time, presence, immersion and engagement, with experiment condition as the only independent variable. Bonferroni corrections were used for all pairwise comparisons. Independent-sample t-tests were used to examine the impact of other independent variables (trial order and height) on user experience ratings and metrics.

\begin{figure}
 \centering
 \includegraphics[width=\columnwidth]{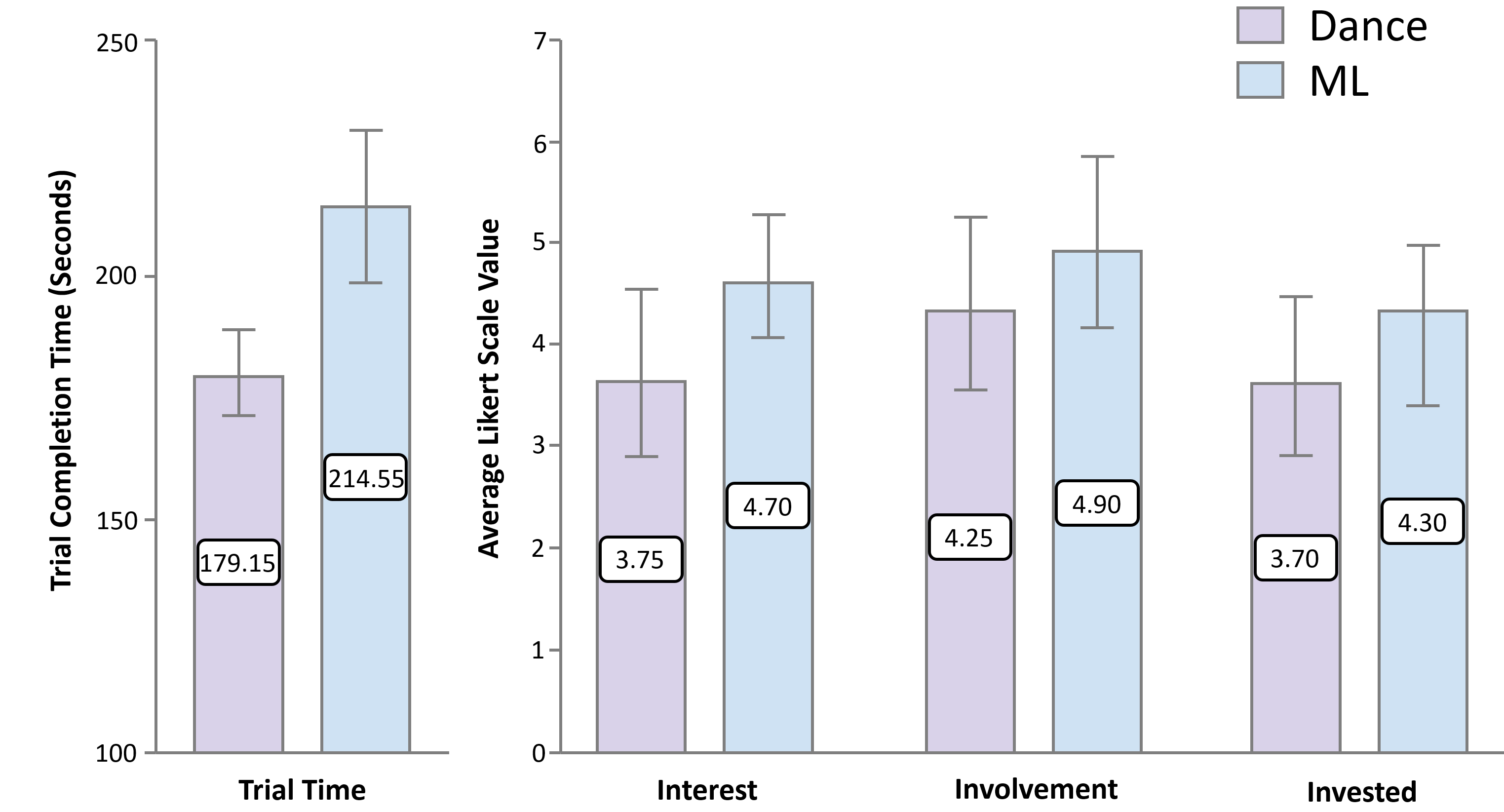}
 \caption{Trial completion time (left) and user experience ratings (right) for the two experiment conditions. Participants spent more time in the experience in the ML condition and also reported higher interest, involvement (\textit{``How much did the auditory aspects of the environment involve you?"}), and investment(\textit{``I was so involved that I felt that my actions could affect the activity"}). For this and all following charts: Error bars represent 95\% confidence interval. User experience ratings were on a 7-point Likert scale (7=Strongly agree/Very much, 1=Strongly disagree/Not at all).}
 \label{fig:audience-type}
\end{figure}

\section{Main Study Results}

We analyzed data to compare the two experiment conditions and investigated the effect of trial order and participant height on reactions to the experience.

\subsection{Audience Type}
We compared the two main experiment conditions--virtual audience present (ML) and no virtual audience present (Dance)--using both qualitative metrics of user experience (interest and involvement) as well as a quantitative metric of engagement (time spent in the experience). Results are highlighted in Figure \ref{fig:audience-type}.

A one-way repeated-measures ANOVA indicated a significant difference in interest between the two conditions ($F_{(1,19)}=4.524, p=.047, \eta_{p}^{2}=.192$), with the post-hoc test showing that participants reported more interest in the experience during the ML condition compared to Dance ($p<.05$). There was also a significant difference in involvement ($F_{(1,19)}=6.011, p=.024, \eta_{p}^{2}=.240$), with participants reporting more involvement with the auditory aspects of the experience (\textit{``How much did the auditory aspects of the environment involve you?"}) in the ML condition ($p<.05$). Participants felt that their actions could influence the experience differently ($F_{(1,19)}=4.557, p=.046, \eta_{p}^{2}=.193$), with a stronger perceived influence in ML compared to Dance ($p<.05$).

Time spent in the experience changed as a function of the experiment condition ($F_{(1,19)}=28.672, p<.001, \eta_{p}^{2}=.601$). Participants spent more time watching the performance in the ML condition compared to Dance ($p<.001$), which tracks with the increased interest score from GEQ questions~\cite{geq2013} in this condition. These results suggest that virtual audiences had a positive impact on user experience and engagement in shared immersive performances.

\subsection{Trial Order}

\begin{figure}
 \centering
 \includegraphics[width=\columnwidth]{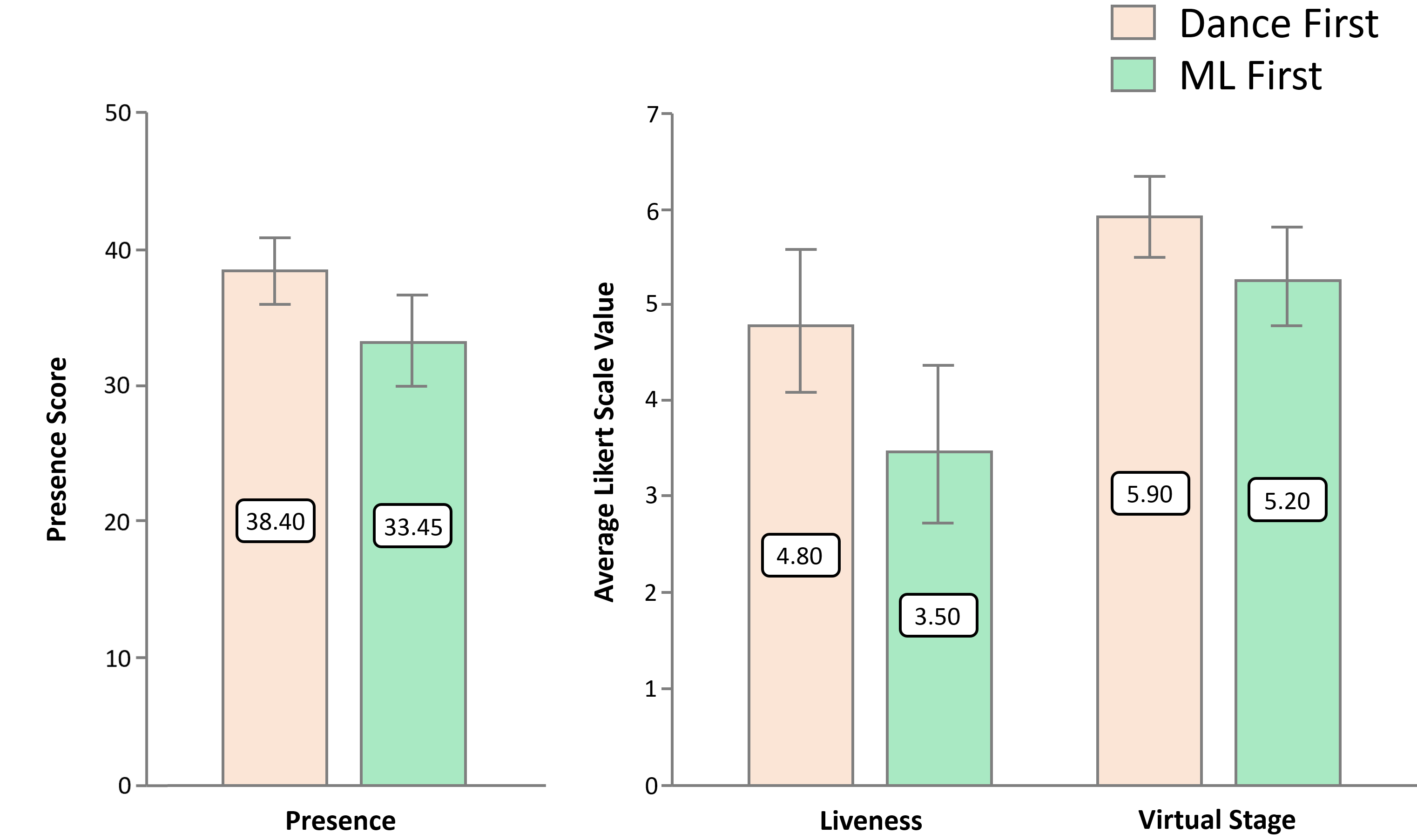}
 \caption{Presence (left) and user experience ratings (right) for the two groups of participants based on trial order (which condition they experienced first). Participants who experienced the Dance (no audience) condition first had a higher presence score, higher liveness (feeling of actually being in a live performance), and appreciated the virtual stage more.}
 \label{fig:trial-order}
\end{figure}

When comparing participants' reactions to their entire study experience (recorded after both sessions, and regarding their impressions of both sessions combined), participants who saw the Dance session first, followed by ML, had a more positive reaction to the experience than participants who saw the ML session first (Figure \ref{fig:trial-order}). This was demonstrated by a higher presence ($t(38)=2.145, p<.05$), higher appreciation of the virtual stage design ($t(38)=1.807, p<.05$), and a stronger feeling of actually being in a live performance ($t(38)=2.145, p<.05$). These results suggest that the order of introduction to the virtual audience could alter user experience and appreciation of the stage and performance.
% -------------------------------------------------------------------------------------------------------------------------------------------------

\begin{figure*}
\centering
  \includegraphics[width=1.0 \textwidth]{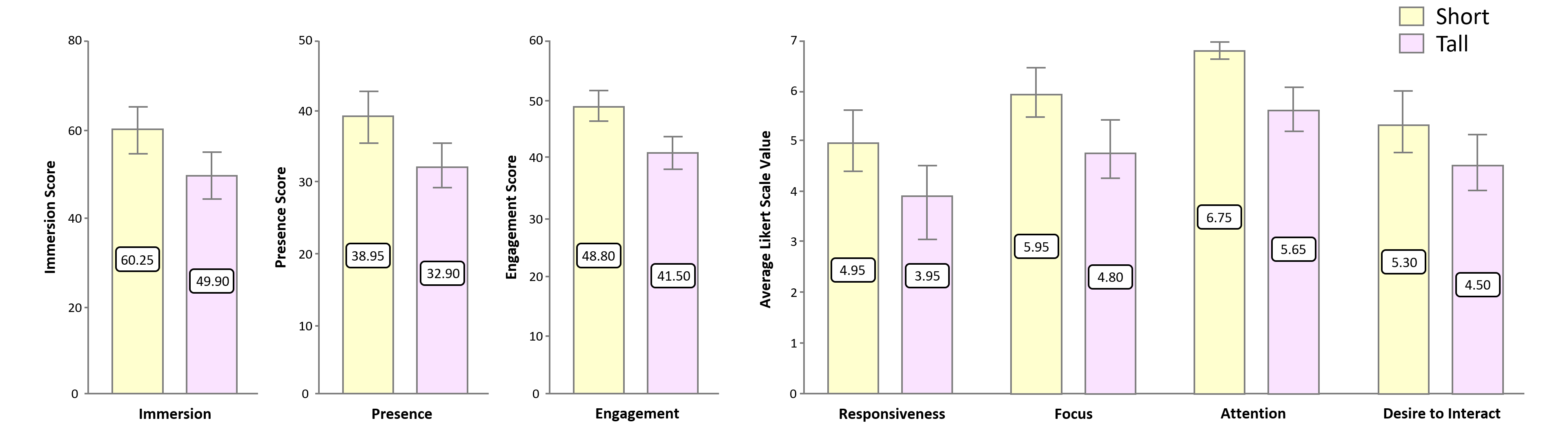}
\caption{Immersion, presence, engagement, and user experience ratings for the two height groups. Shorter participants had higher immersion, presence and engagement. They also reported more focus, more attention, more of a desire to interact with the virtual characters and felt that the environment was more responsive to their actions.}
 \label{fig:height}
\end{figure*}

\subsection{Height}
Considering how height influences a person's live concert experience, we investigated the role of height and how it affects the desire to watch an AR play catered to a freely walking user. Therefore, we divided the 20 participants into two groups of equal size based on their height--the 10 tallest participants (\textit{Tall}) in one group and the 10 shortest participants in another (\textit{Short}). Results are highlighted in Figure \ref{fig:height}. An independent-samples t-test indicated that the shorter participants had significantly higher immersion ($t(38)=-3.113, p<.005$), overall presence $t(38)=-3.234, p<.005$) and engagement ($t(38)=-3.799, p<.001$) than taller participants during the study. Shorter participants also felt that the environment was more responsive to their actions ($t(38)=-2.380, p=.022$), were more focused on the activity ($t(38)=-3.598, p=.001$), and wanted to interact with the virtual characters and objects more ($t(38)=-2.312, p=.026$) than taller participants. 

On the other hand, taller participants felt like they had more external distractions than shorter participants ($t(38)=-2.427, p=.020$), and also felt that the virtual content captured their attention less than the shorter participants did($t(38)=-4.341, p<.001$).To verify our results, we compared the 5 tallest participants' and the 5 shortest participants' experiences. Our results held, with even stronger significance being demonstrated. An examination of the participant's gender did not result in significant variations. The observation that shorter participants enjoyed the experience more, irrespective of gender, indicates that this effect cannot be attributed to gender differences alone. Additionally, we did not find any other correlations that could explain this effect.

%%This analysis indicates that shorter participants appeared to have a more positive reaction to the MR experience than taller participants did. This is interesting as the shorter group was, on average, similar in height(165.3$cm$) to the female dancers and audience avatars (164$cm$), and shorter than the male audience avatars (176$cm$). Conversely, the taller participant group (average height of 179.4 $cm$) was taller than both female and male virtual audience avatars in the play.

% -------------------------------------------------------------------------------------------------------------------------------------------------

\section{Discussion}

Audience Amplified explored the inclusion of virtual audiences in AR theater applications and their impact on user experience. We introduce a training combining imitation and reinforcement learning to create virtual audiences for target space layout that accurately capture the nuances of human imperfections and behaviors.

Participants had increased involvement, interest, and engagement with the experience when virtual audiences were present. There appear to be a number of reasons for this, including the experience being more realistic (\textit{``It definitely brought more of a real experience compared to having no audience at all."}, P7) and social (\textit{``I think that the social aspects are the best part, because ... digesting (the experience)) together makes the experience meaningful..."}, P3; \textit{``it truly felt like you were interacting with others; more so than an online video game"}, P10). Some participants mentioned, however, that the virtual audience sometimes blocked their view of the dancers. This might have happened as a result of the virtual audience being trained using data gathered from an earlier study \cite{kim2023dynamic}, which made use of user data from encounters with AR Theater alone and without audiences. 

Additionally, participants enjoyed the experience more when interacting with the ML-trained virtual audience, feeling higher levels of presence and appreciation for the content. The auditory aspect enhanced their sense of environment and engagement, especially when they first experienced the play without the virtual audience and were introduced to it later.

Participants who experienced the virtual audience after initially watching the performance with no audience expressed a greater appreciation for the dance play and stage environment, as they were able to enjoy the play without any interruptions. While virtual audiences can enhance presence and enjoyment, there is a tradeoff to consider. This insight is crucial for making design choices in AR theater, as adding an audience does not always enhance the experience; it depends on the director's objectives. We believe that using a combination of scenes, sometimes with an audience and other times in solitude, can be a powerful tool for AR theater.

Shorter participants reported a more positive experience overall than taller participants, with higher reported presence, immersion, and engagement. Taller participants also reported more external distractions and less focus on the activity, which could be driven by the fact that all of the virtual characters were much shorter than the taller participant group on average while the shorter participant group was about the same height as, or much shorter than, the virtual avatars. For example, a participant from the taller group said \textit{``What drew me out were ... the different sizes of the avatars"} (P19). Further exploration is needed to determine the impact of scale and content placement on user experience in immersive storytelling, but our results suggest that it is important to consider user height and perspective when designing such experiences.

\section{Limitations and Future Work}

While Audience Amplified offers the sensation of being in the company of others, it is important to note that users are, in reality, engaging with these virtual audiences alone. Not directly comparing the experience with a real audience is a limitation.  Future research should examine how virtual audiences perform in group settings and how this alters the user experience. Despite their realistic appearance and behavior, our virtual audiences are not interactive. Furthermore, compared to the NPC model used, our virtual audience exhibited more distributed attention and did not provide full focus on the performance. Therefore, we aim to refine our avatar model to enhance real-time content engagement in future projects. Surprisingly, some participants spontaneously danced along with the avatar dancer, suggesting that virtual audiences perhaps should at times react to and participate in unexpected situations. Finally, While our virtual audience can be implemented in various environments, their behavior was most effective in the corridor layout we tested. Future projects will investigate training virtual avatars to adapt to building data and user context, using deep neural networks to automatically adjust to floor layouts.

\section{Conclusion}
We present Audience Amplified, a novel augmented reality dance performance that includes trained virtual audience avatars that perform human-like behaviors and movements using imitation learning. Two pilot studies were held to determine the optimal appearance and behavior of the audience avatars. A following 20-subject user study was conducted to understand differences in user experience when viewing the performance with and without the virtual audience. We found that participants reacted positively to the experience that included virtual audiences and appreciated the social aspect of watching a performance with other virtual audiences, especially as it provided live-event like experience. 

Some important design considerations for mixed reality theatrical experiences are highlighted in our analysis, including the presence and absence of virtual audiences in the interactive narrative, as well as optimizing content placement and scale based on user height and perspective. 

In this project, we focused on experiences that followed the participatory open-world exploration style, which provided users freedom to personalize their experience as a viewer. We see great possibilities for AR performances that can be experienced via mobile AR by individual audience members at their leisure, while still experiencing some semblance of audience connection. In the future, the AR spectator’s social viewing behavior could influence the amplified audience experience of subsequent users, leading to a new kind of asynchronous, yet still, socially connected experience.

\acknowledgments{%
	This work was supported in part by NSF award IIS-2211784. The authors thank Kangyou Yu and Radha Kumaran for their assistance and friendship throughout this project.%
}

\bibliographystyle{abbrv-doi}

\bibliography{template}

\begin{thebibliography}{10}

\bibitem{ALAMOODI2021114155}
A.~Alamoodi, B.~Zaidan, A.~Zaidan, O.~Albahri, K.~Mohammed, R.~Malik, E.~Almahdi, M.~Chyad, Z.~Tareq, A.~Albahri, H.~Hameed, and M.~Alaa.
\newblock Sentiment analysis and its applications in fighting covid-19 and infectious diseases: A systematic review.
\newblock {\em Expert Systems with Applications}, 167:114155, 2021. doi: {{%
10\hspace{.1pt}\discretionary{.}{%
}{.}\hspace{.4pt}1016\discretionary{/}{%
}{/}j\hspace{.1pt}\discretionary{.}{%
}{.}\hspace{.4pt}eswa\hspace{.1pt}\discretionary{.}{%
}{.}\hspace{.4pt}2020\hspace{.1pt}\discretionary{.}{%
}{.}\hspace{.4pt}114155}}


\bibitem{10.1145/985921.986048}
E.~Brown and P.~Cairns.
\newblock A grounded investigation of game immersion.
\newblock In {\em CHI '04 Extended Abstracts on Human Factors in Computing Systems}, CHI EA '04, p. 1297–1300. Association for Computing Machinery, New York, NY, USA, 2004. doi: {{%
10\hspace{.1pt}\discretionary{.}{%
}{.}\hspace{.4pt}1145\discretionary{/}{%
}{/}985921\hspace{.1pt}\discretionary{.}{%
}{.}\hspace{.4pt}986048}}


\bibitem{buck2022azure}
A.~Buck, R.~Parker, and M.~Wojciakowski.
\newblock Azure spatial anchors overview.
\newblock https://learn.microsoft.com/en-us/azure/spatial-anchors/overview, 2022.
\newblock Accessed: October 2022.

\bibitem{campo2023assessment}
A.~Campo, A.~Michałko, B.~{Van Kerrebroeck}, B.~Stajic, M.~Pokric, and M.~Leman.
\newblock The assessment of presence and performance in an ar environment for motor imitation learning: A case-study on violinists.
\newblock {\em Computers in Human Behavior}, 146:107810, 2023. doi: {{%
10\hspace{.1pt}\discretionary{.}{%
}{.}\hspace{.4pt}1016\discretionary{/}{%
}{/}j\hspace{.1pt}\discretionary{.}{%
}{.}\hspace{.4pt}chb\hspace{.1pt}\discretionary{.}{%
}{.}\hspace{.4pt}2023\hspace{.1pt}\discretionary{.}{%
}{.}\hspace{.4pt}107810}}


\bibitem{game2020capsule}
CAPSULEOFFICIAL.
\newblock Capsule house - public, 2020.

\bibitem{cheok2002interactive}
A.~Cheok, W.~Weihua, X.~Yang, S.~Prince, F.~S. Wan, M.~Billinghurst, and H.~Kato.
\newblock Interactive theatre experience in embodied + wearable mixed reality space.
\newblock In {\em Proceedings. International Symposium on Mixed and Augmented Reality}, pp. 59--317, 2002. doi: {{%
10\hspace{.1pt}\discretionary{.}{%
}{.}\hspace{.4pt}1109\discretionary{/}{%
}{/}ISMAR\hspace{.1pt}\discretionary{.}{%
}{.}\hspace{.4pt}2002\hspace{.1pt}\discretionary{.}{%
}{.}\hspace{.4pt}1115073}}


\bibitem{corbin2014basics}
J.~Corbin and A.~Strauss.
\newblock {\em Basics of qualitative research: Techniques and procedures for developing grounded theory}.
\newblock Sage publications, fourth ed., 2014.

\bibitem{coulombe2021virtual}
A.~Coulombe, D.~Gochfeld, B.~Bradley, K.~Laibson, R.~Long, and R.~Miletitch.
\newblock Virtual reality live theatre on no budget: A model for independent theatrical productions using open-source social vr.
\newblock In {\em ACM SIGGRAPH 2021 Educators Forum}, SIGGRAPH '21. Association for Computing Machinery, New York, NY, USA, 2021. doi: {{%
10\hspace{.1pt}\discretionary{.}{%
}{.}\hspace{.4pt}1145\discretionary{/}{%
}{/}3450549\hspace{.1pt}\discretionary{.}{%
}{.}\hspace{.4pt}3464413}}


\bibitem{dow2005exploring}
S.~Dow, J.~Lee, C.~Oezbek, B.~Maclntyre, J.~D. Bolter, and M.~Gandy.
\newblock Exploring spatial narratives and mixed reality experiences in oakland cemetery.
\newblock In {\em Proceedings of the 2005 ACM SIGCHI International Conference on Advances in Computer Entertainment Technology}, ACE '05, p. 51–60. Association for Computing Machinery, New York, NY, USA, 2005. doi: {{%
10\hspace{.1pt}\discretionary{.}{%
}{.}\hspace{.4pt}1145\discretionary{/}{%
}{/}1178477\hspace{.1pt}\discretionary{.}{%
}{.}\hspace{.4pt}1178484}}


\bibitem{10.1145/2967934.2968092}
K.~Emmerich and M.~Masuch.
\newblock The influence of virtual agents on player experience and performance.
\newblock In {\em Proceedings of the 2016 Annual Symposium on Computer-Human Interaction in Play}, CHI PLAY '16, p. 10–21. Association for Computing Machinery, New York, NY, USA, 2016. doi: {{%
10\hspace{.1pt}\discretionary{.}{%
}{.}\hspace{.4pt}1145\discretionary{/}{%
}{/}2967934\hspace{.1pt}\discretionary{.}{%
}{.}\hspace{.4pt}2968092}}


\bibitem{watchmeplay}
K.~Emmerich and M.~Masuch.
\newblock {\em Watch Me Play: Does Social Facilitation Apply to Digital Games?}, p.~12.
\newblock Association for Computing Machinery, New York, NY, USA, 2018.

\bibitem{feiner1997touring}
S.~Feiner, B.~MacIntyre, T.~H{\"o}llerer, and A.~Webster.
\newblock A touring machine: Prototyping 3d mobile augmented reality systems for exploring the urban environment.
\newblock {\em Personal Technologies}, 1(4):208--217, 1997.

\bibitem{feuerriegel2018package}
S.~Feuerriegel, N.~Proellochs, and M.~S. Feuerriegel.
\newblock Package ‘sentimentanalysis’.
\newblock {\em CRAN: London, UK}, 2018.

\bibitem{10.1162/pres.1992.1.4.482}
G.~Fontaine.
\newblock {The Experience of a Sense of Presence in Intercultural and International Encounters}.
\newblock {\em Presence: Teleoperators and Virtual Environments}, 1(4):482--490, 11 1992. doi: {{%
10\hspace{.1pt}\discretionary{.}{%
}{.}\hspace{.4pt}1162\discretionary{/}{%
}{/}pres\hspace{.1pt}\discretionary{.}{%
}{.}\hspace{.4pt}1992\hspace{.1pt}\discretionary{.}{%
}{.}\hspace{.4pt}1\hspace{.1pt}\discretionary{.}{%
}{.}\hspace{.4pt}4\hspace{.1pt}\discretionary{.}{%
}{.}\hspace{.4pt}482}}


\bibitem{fujihata2022behere}
M.~Fujihata.
\newblock {{BeHere}} 1942, {{Japanese American National Museum}}.
\newblock https://www.janm.org/exhibits/behere1942, May 2022.

\bibitem{GEORGIOU201724}
Y.~Georgiou and E.~A. Kyza.
\newblock The development and validation of the ari questionnaire: An instrument for measuring immersion in location-based augmented reality settings.
\newblock {\em International Journal of Human-Computer Studies}, 98:24--37, 2017. doi: {{%
10\hspace{.1pt}\discretionary{.}{%
}{.}\hspace{.4pt}1016\discretionary{/}{%
}{/}j\hspace{.1pt}\discretionary{.}{%
}{.}\hspace{.4pt}ijhcs\hspace{.1pt}\discretionary{.}{%
}{.}\hspace{.4pt}2016\hspace{.1pt}\discretionary{.}{%
}{.}\hspace{.4pt}09\hspace{.1pt}\discretionary{.}{%
}{.}\hspace{.4pt}014}}


\bibitem{gochfeld2018holojam}
D.~Gochfeld, C.~Brenner, K.~Layng, S.~Herscher, C.~DeFanti, M.~Olko, D.~Shinn, S.~Riggs, C.~Fern\'{a}ndez-Vara, and K.~Perlin.
\newblock Holojam in wonderland: Immersive mixed reality theater.
\newblock In {\em ACM SIGGRAPH 2018 Art Gallery}, SIGGRAPH '18, p. 362–367. Association for Computing Machinery, New York, NY, USA, 2018. doi: {{%
10\hspace{.1pt}\discretionary{.}{%
}{.}\hspace{.4pt}1145\discretionary{/}{%
}{/}3202918\hspace{.1pt}\discretionary{.}{%
}{.}\hspace{.4pt}3203091}}


\bibitem{VR_SharedSocialSpaces}
J.~Gugenheimer, C.~Mai, M.~McGill, J.~Williamson, F.~Steinicke, and K.~Perlin.
\newblock Challenges using head-mounted displays in shared and social spaces.
\newblock In {\em Extended Abstracts of the 2019 CHI Conference on Human Factors in Computing Systems}, CHI EA '19, p. 1–8. Association for Computing Machinery, New York, NY, USA, 2019. doi: {{%
10\hspace{.1pt}\discretionary{.}{%
}{.}\hspace{.4pt}1145\discretionary{/}{%
}{/}3290607\hspace{.1pt}\discretionary{.}{%
}{.}\hspace{.4pt}3299028}}


\bibitem{10.1145/3290688.3290752}
J.~C. Haller, Y.~H. Jang, J.~Haller, L.~Shaw, and B.~C. W\"{u}nsche.
\newblock Hiit the road: Using virtual spectator feedback in hiit-based exergaming.
\newblock In {\em Proceedings of the Australasian Computer Science Week Multiconference}, ACSW 2019. Association for Computing Machinery, New York, NY, USA, 2019. doi: {{%
10\hspace{.1pt}\discretionary{.}{%
}{.}\hspace{.4pt}1145\discretionary{/}{%
}{/}3290688\hspace{.1pt}\discretionary{.}{%
}{.}\hspace{.4pt}3290752}}


\bibitem{article2021npc}
D.~Heaney.
\newblock This openai gpt-3 powered demo is a glimpse of npcs in the future, 2021.

\bibitem{herscher2019cavrn}
S.~Herscher, C.~DeFanti, N.~G. Vitovitch, C.~Brenner, H.~Xia, K.~Layng, and K.~Perlin.
\newblock Cavrn: an exploration and evaluation of a collective audience virtual reality nexus experience.
\newblock In {\em Proceedings of the 32nd Annual ACM Symposium on User Interface Software and Technology}, pp. 1137--1150, 2019.

\bibitem{ho2022perspective}
J.~C. Ho and R.~Ng.
\newblock Perspective-taking of non-player characters in prosocial virtual reality games: Effects on closeness, empathy, and game immersion.
\newblock {\em Behaviour \& Information Technology}, 41(6):1185--1198, 2022.

\bibitem{game2021murder}
Jar.
\newblock Murder 4 - {{Worlds}} on {{VRChat}}, 2021.

\bibitem{kao_effects_2021}
D.~Kao.
\newblock The effects of observation in video games: how remote observation influences player experience, motivation, and behaviour.
\newblock {\em Behaviour \& Information Technology}, pp. 1--23, Apr. 2021.
\newblock Publisher: Taylor \& Francis. doi: {{%
10\hspace{.1pt}\discretionary{.}{%
}{.}\hspace{.4pt}1080\discretionary{/}{%
}{/}0144929X\hspace{.1pt}\discretionary{.}{%
}{.}\hspace{.4pt}2021\hspace{.1pt}\discretionary{.}{%
}{.}\hspace{.4pt}1906321}}


\bibitem{kim2023dynamic}
Y.-J. Kim, J.~Lu, and T.~Höllerer.
\newblock Dynamic theater: Location-based immersive dance theater, investigating user guidance and experience.
\newblock In {\em Proceedings of the 29th ACM Symposium on Virtual Reality Software and Technology}, VRST '23. Association for Computing Machinery, New York, NY, USA, 2023. doi: {{%
10\hspace{.1pt}\discretionary{.}{%
}{.}\hspace{.4pt}1145\discretionary{/}{%
}{/}3611659\hspace{.1pt}\discretionary{.}{%
}{.}\hspace{.4pt}3615705}}


\bibitem{kim2023rdr}
Y.-J. Kim, A.~D. Wilson, J.~Jacobs, and T.~Höllerer.
\newblock Reality distortion room: A study of user locomotion responses to spatial augmented reality effects.
\newblock In {\em 2023 IEEE International Symposium on Mixed and Augmented Reality (ISMAR)}, pp. 1201--1210, 2023. doi: {{%
10\hspace{.1pt}\discretionary{.}{%
}{.}\hspace{.4pt}1109\discretionary{/}{%
}{/}ISMAR59233\hspace{.1pt}\discretionary{.}{%
}{.}\hspace{.4pt}2023\hspace{.1pt}\discretionary{.}{%
}{.}\hspace{.4pt}00137}}


\bibitem{kisker2021behavioral}
J.~Kisker, T.~Gruber, and B.~Sch{\"o}ne.
\newblock Behavioral realism and lifelike psychophysiological responses in virtual reality by the example of a height exposure.
\newblock {\em Psychological research}, 85:68--81, 2021.

\bibitem{CAVE_Leon}
K.~Layng, K.~Perlin, S.~Herscher, C.~Brenner, and T.~Meduri.
\newblock {CAVE: Making Collective Virtual Narrative: Best Paper Award}.
\newblock {\em Leonardo}, 52(4):349--356, 08 2019. doi: {{%
10\hspace{.1pt}\discretionary{.}{%
}{.}\hspace{.4pt}1162\discretionary{/}{%
}{/}leon\_a\_01776}}


\bibitem{lessiter2001cross}
J.~Lessiter, J.~Freeman, E.~Keogh, and J.~Davidoff.
\newblock A cross-media presence questionnaire: The itc-sense of presence inventory.
\newblock {\em Presence: Teleoperators \& Virtual Environments}, 10(3):282--297, 2001.

\bibitem{10.1145/2077451.2077464}
M.~Leyrer, S.~A. Linkenauger, H.~H. B\"{u}lthoff, U.~Kloos, and B.~Mohler.
\newblock The influence of eye height and avatars on egocentric distance estimates in immersive virtual environments.
\newblock In {\em Proceedings of the ACM SIGGRAPH Symposium on Applied Perception in Graphics and Visualization}, APGV '11, p. 67–74. Association for Computing Machinery, New York, NY, USA, 2011. doi: {{%
10\hspace{.1pt}\discretionary{.}{%
}{.}\hspace{.4pt}1145\discretionary{/}{%
}{/}2077451\hspace{.1pt}\discretionary{.}{%
}{.}\hspace{.4pt}2077464}}


\bibitem{lyons2023gumball}
D.~V. Lyons, C.~L. Davis, S.~Butchko, W.~Frank, B.~Tull, and B.~Roy.
\newblock Gumball dreams: Live theatre in vr.
\newblock In {\em ACM SIGGRAPH 2023 Immersive Pavilion}, SIGGRAPH '23. Association for Computing Machinery, New York, NY, USA, 2023. doi: {{%
10\hspace{.1pt}\discretionary{.}{%
}{.}\hspace{.4pt}1145\discretionary{/}{%
}{/}3588027\hspace{.1pt}\discretionary{.}{%
}{.}\hspace{.4pt}3595593}}


\bibitem{macey2023feeling}
A.-L. Macey, S.~J{\"a}rvel{\"a}, D.~Fernandez~Galeote, and J.~Hamari.
\newblock Feeling small or standing tall? height manipulation affects speech anxiety and arousal in virtual reality.
\newblock {\em Cyberpsychology, Behavior, and Social Networking}, 26(4):246--254, 2023.

\bibitem{madsen2022fear}
P.~Madsen, H.~Pohl, and T.~Merritt.
\newblock Fear inducing play in an ar escape room with human and robotic npcs.
\newblock In {\em Extended Abstracts of the 2022 Annual Symposium on Computer-Human Interaction in Play}, pp. 38--43, 2022.

\bibitem{meehan2002physiological}
M.~Meehan, B.~Insko, M.~Whitton, and F.~P. Brooks~Jr.
\newblock Physiological measures of presence in stressful virtual environments.
\newblock {\em Acm transactions on graphics (tog)}, 21(3):645--652, 2002.

\bibitem{merrick2006motivated}
K.~Merrick and M.~L. Maher.
\newblock Motivated reinforcement learning for non-player characters in persistent computer game worlds.
\newblock In {\em Proceedings of the 2006 ACM SIGCHI International Conference on Advances in Computer Entertainment Technology}, ACE '06, p. 3–es. Association for Computing Machinery, New York, NY, USA, 2006. doi: {{%
10\hspace{.1pt}\discretionary{.}{%
}{.}\hspace{.4pt}1145\discretionary{/}{%
}{/}1178823\hspace{.1pt}\discretionary{.}{%
}{.}\hspace{.4pt}1178828}}


\bibitem{merrick2007reinforcement}
K.~E. Merrick and M.~L. Maher.
\newblock Motivated reinforcement learning for adaptive characters in open-ended simulation games.
\newblock ACE '07, p. 127–134. Association for Computing Machinery, New York, NY, USA, 2007. doi: {{%
10\hspace{.1pt}\discretionary{.}{%
}{.}\hspace{.4pt}1145\discretionary{/}{%
}{/}1255047\hspace{.1pt}\discretionary{.}{%
}{.}\hspace{.4pt}1255073}}


\bibitem{ARGulliver}
G.~Mileva.
\newblock Traditional theater gets an augmented reality makeover.
\newblock https://arpost.co/2021/02/24/traditional-theater-augmented-reality/, February 2021.

\bibitem{nicholas2021expanding}
M.~J. Nicholas, S.~C. Daffara, and E.~Paulos.
\newblock Expanding the design space for technology-mediated theatre experiences.
\newblock In {\em Designing Interactive Systems Conference 2021}, pp. 2026--2038, 2021.

\bibitem{geq2013}
K.~L. Norman.
\newblock {GEQ (Game Engagement/Experience Questionnaire): A Review of Two Papers}.
\newblock {\em Interacting with Computers}, 25(4):278--283, 03 2013. doi: {{%
10\hspace{.1pt}\discretionary{.}{%
}{.}\hspace{.4pt}1093\discretionary{/}{%
}{/}iwc\discretionary{/}{%
}{/}iwt009}}


\bibitem{game2021shrine}
T.~Null.
\newblock Sun shower shrine sanctuary - worlds on vrchat, 2021.

\bibitem{Play_Leon}
N.~O’Dwyer, N.~Johnson, E.~Bates, R.~Pagés, J.~Ondřej, K.~Amplianitis, D.~Monaghan, and A.~Smolic.
\newblock {Samuel Beckett in Virtual Reality: Exploring Narrative Using Free Viewpoint Video}.
\newblock {\em Leonardo}, 54(2):166--171, 04 2021. doi: {{%
10\hspace{.1pt}\discretionary{.}{%
}{.}\hspace{.4pt}1162\discretionary{/}{%
}{/}leon\_a\_01721}}


\bibitem{doi:10.1177/0018720816669646}
M.~H. Phan, J.~R. Keebler, and B.~S. Chaparro.
\newblock The development and validation of the game user experience satisfaction scale (guess).
\newblock {\em Human Factors}, 58(8):1217--1247, 2016.
\newblock PMID: 27647156. doi: {{%
10\hspace{.1pt}\discretionary{.}{%
}{.}\hspace{.4pt}1177\discretionary{/}{%
}{/}0018720816669646}}


\bibitem{pietroszek2022dill}
K.~Pietroszek, M.~Rebol, and B.~Lake.
\newblock Dill pickle: Interactive theatre play in virtual reality.
\newblock In {\em Proceedings of the 28th ACM Symposium on Virtual Reality Software and Technology}, pp. 1--2, 2022.

\bibitem{pietroszek2022meeting}
K.~Pietroszek, M.~Rebol, and B.~Lake.
\newblock The meeting: Volumetric participatory theatre play in mixed reality.
\newblock In {\em Proceedings of the 10th International Conference on Human-Agent Interaction}, pp. 330--332, 2022.

\bibitem{piumsomboon2018minime}
T.~Piumsomboon, G.~A. Lee, J.~D. Hart, B.~Ens, R.~W. Lindeman, B.~H. Thomas, and M.~Billinghurst.
\newblock Mini-me: An adaptive avatar for mixed reality remote collaboration.
\newblock In {\em Proceedings of the 2018 CHI Conference on Human Factors in Computing Systems}, CHI '18, p. 1–13. Association for Computing Machinery, New York, NY, USA, 2018. doi: {{%
10\hspace{.1pt}\discretionary{.}{%
}{.}\hspace{.4pt}1145\discretionary{/}{%
}{/}3173574\hspace{.1pt}\discretionary{.}{%
}{.}\hspace{.4pt}3173620}}


\bibitem{web2023reddit}
{Reddit: r/VRchat}.
\newblock {VR Chat Worlds You Should Visit by TheVrHub}, 2023.

\bibitem{rompapas2018holoroyale}
D.~Rompapas, C.~Sandor, A.~Plopski, D.~Saakes, D.~H. Yun, T.~Taketomi, and H.~Kato.
\newblock Holoroyale: A large scale high fidelity augmented reality game.
\newblock In {\em Adjunct Proceedings of the 31st Annual ACM Symposium on User Interface Software and Technology}, pp. 163--165, 2018.

\bibitem{slater2003note}
M.~Slater.
\newblock A note on presence terminology.
\newblock {\em Presence connect}, 3(3):1--5, 2003.

\bibitem{slater2023sentiment}
M.~Slater, C.~Cabriera, G.~Senel, D.~Banakou, A.~Beacco, R.~Oliva, and J.~Gallego.
\newblock The sentiment of a virtual rock concert.
\newblock {\em Virtual Reality}, 27(2):651--675, 2023.

\bibitem{mel2022sentiment}
M.~Slater, C.~Cabriera, G.~Şenel, D.~Banakou, A.~Beacco, R.~Oliva, and J.~Gallego.
\newblock The sentiment of a virtual rock concert.
\newblock {\em Virtual Reality}, 27:1--25, 08 2022. doi: {{%
10\hspace{.1pt}\discretionary{.}{%
}{.}\hspace{.4pt}1007\discretionary{/}{%
}{/}s10055\discretionary{%
}{-}{-}022\discretionary{%
}{-}{-}00685\discretionary{%
}{-}{-}9}}


\bibitem{doi:10.1080/10447318.2022.2121038}
S.~W. Song and M.~Shin.
\newblock Uncanny valley effects on chatbot trust, purchase intention, and adoption intention in the context of e-commerce: The moderating role of avatar familiarity.
\newblock {\em International Journal of Human–Computer Interaction}, 0(0):1--16, 2022. doi: {{%
10\hspace{.1pt}\discretionary{.}{%
}{.}\hspace{.4pt}1080\discretionary{/}{%
}{/}10447318\hspace{.1pt}\discretionary{.}{%
}{.}\hspace{.4pt}2022\hspace{.1pt}\discretionary{.}{%
}{.}\hspace{.4pt}2121038}}


\bibitem{game2019angelica}
O.~S. Studio.
\newblock Dear {{Angelica}} on {{Oculus Rift}}, 2019.

\bibitem{10.1145/2927929.2927955}
K.~Tcha-Tokey, E.~Loup-Escande, O.~Christmann, and S.~Richir.
\newblock A questionnaire to measure the user experience in immersive virtual environments.
\newblock In {\em Proceedings of the 2016 Virtual Reality International Conference}, VRIC '16. Association for Computing Machinery, New York, NY, USA, 2016. doi: {{%
10\hspace{.1pt}\discretionary{.}{%
}{.}\hspace{.4pt}1145\discretionary{/}{%
}{/}2927929\hspace{.1pt}\discretionary{.}{%
}{.}\hspace{.4pt}2927955}}


\bibitem{game2019presents}
{Tender Claws}.
\newblock {The Under Presents}, 2019.

\bibitem{thomas2000arquake}
B.~Thomas, B.~Close, J.~Donoghue, J.~Squires, P.~De~Bondi, M.~Morris, and W.~Piekarski.
\newblock Arquake: An outdoor/indoor augmented reality first person application.
\newblock In {\em Digest of papers. Fourth international symposium on wearable computers}, pp. 139--146. IEEE, 2000.

\bibitem{trivedi2023human}
H.~Trivedi and C.~Mousas.
\newblock Human-virtual crowd interaction: Towards understanding the effects of crowd avoidance proximity in an immersive virtual environment.
\newblock {\em Computer Animation and Virtual Worlds}, 34(3-4):e2169, 2023.

\bibitem{usoh2000using}
M.~Usoh, E.~Catena, S.~Arman, and M.~Slater.
\newblock Using presence questionnaires in reality.
\newblock {\em Presence}, 9(5):497--503, 2000.

\bibitem{web2023mocap}
{Vicon}.
\newblock {Vicon Motion Capture Cameras System}, 2023.

\bibitem{game2023japan}
{Virtual Remix Japan}.
\newblock Live vr japanese concert - v-kitazawa awake in vrchat livestream, 2023.

\bibitem{wang2020effect}
T.-Y. Wang, Y.~Sato, M.~Otsuki, H.~Kuzuoka, and Y.~Suzuki.
\newblock Effect of body representation level of an avatar on quality of ar-based remote instruction.
\newblock {\em Multimodal Technologies and Interaction}, 4(1):3, 2020.

\bibitem{wang2022exploring}
Z.~Wang and Y.-J. Kim.
\newblock Exploring immersive mixed reality simulations and its impact on climate change awareness.
\newblock {\em Asian Journal of Applied Science and Engineering}, 11(1):1--6, 2022.

\bibitem{weng2011soul}
D.~Weng, W.~Xu, D.~Li, Y.~Wang, and Y.~Liu.
\newblock “soul hunter”: A novel augmented reality application in theme parks.
\newblock In {\em 2011 10th IEEE International Symposium on Mixed and Augmented Reality}, pp. 279--280. IEEE, 2011.

\bibitem{10.1162/105474698565686}
B.~G. Witmer and M.~J. Singer.
\newblock {Measuring Presence in Virtual Environments: A Presence Questionnaire}.
\newblock {\em Presence: Teleoperators and Virtual Environments}, 7(3):225--240, 06 1998. doi: {{%
10\hspace{.1pt}\discretionary{.}{%
}{.}\hspace{.4pt}1162\discretionary{/}{%
}{/}105474698565686}}


\bibitem{Jacob_ART}
J.~O. Wobbrock, L.~Findlater, D.~Gergle, and J.~J. Higgins.
\newblock The aligned rank transform for nonparametric factorial analyses using only anova procedures.
\newblock In {\em Proceedings of the SIGCHI Conference on Human Factors in Computing Systems}, CHI '11, p. 143–146. Association for Computing Machinery, New York, NY, USA, 2011. doi: {{%
10\hspace{.1pt}\discretionary{.}{%
}{.}\hspace{.4pt}1145\discretionary{/}{%
}{/}1978942\hspace{.1pt}\discretionary{.}{%
}{.}\hspace{.4pt}1978963}}


\bibitem{wojtusch2015humod}
J.~Wojtusch and O.~von Stryk.
\newblock Humod - a versatile and open database for the investigation, modeling and simulation of human motion dynamics on actuation level.
\newblock In {\em 2015 IEEE-RAS 15th International Conference on Humanoid Robots (Humanoids)}, pp. 74--79, 2015. doi: {{%
10\hspace{.1pt}\discretionary{.}{%
}{.}\hspace{.4pt}1109\discretionary{/}{%
}{/}HUMANOIDS\hspace{.1pt}\discretionary{.}{%
}{.}\hspace{.4pt}2015\hspace{.1pt}\discretionary{.}{%
}{.}\hspace{.4pt}7363534}}


\bibitem{game2023flow}
xr~marketplace.
\newblock Standard ticket ~ virtual remix japan × flow virtual live, 2023.

\bibitem{10.3389/fpsyg.2023.1079132}
W.~Xu, K.~Yu, X.~Meng, D.~Monteiro, D.~Kao, and H.-N. Liang.
\newblock {Exploring the effect of the Group Size and Feedback of non-player character spectators in virtual reality exergames}.
\newblock {\em Frontiers in Psychology}, 14, 2023. doi: {{%
10\hspace{.1pt}\discretionary{.}{%
}{.}\hspace{.4pt}3389\discretionary{/}{%
}{/}fpsyg\hspace{.1pt}\discretionary{.}{%
}{.}\hspace{.4pt}2023\hspace{.1pt}\discretionary{.}{%
}{.}\hspace{.4pt}1079132}}


\bibitem{yakura2020enhancing}
H.~Yakura and M.~Goto.
\newblock Enhancing participation experience in vr live concerts by improving motions of virtual audience avatars.
\newblock In {\em 2020 IEEE International Symposium on Mixed and Augmented Reality (ISMAR)}, pp. 555--565, 2020. doi: {{%
10\hspace{.1pt}\discretionary{.}{%
}{.}\hspace{.4pt}1109\discretionary{/}{%
}{/}ISMAR50242\hspace{.1pt}\discretionary{.}{%
}{.}\hspace{.4pt}2020\hspace{.1pt}\discretionary{.}{%
}{.}\hspace{.4pt}00083}}


\bibitem{yang2019dreamwalker}
J.~J. Yang, C.~Holz, E.~Ofek, and A.~D. Wilson.
\newblock Dreamwalker: Substituting real-world walking experiences with a virtual reality.
\newblock In {\em Proceedings of the 32nd Annual ACM Symposium on User Interface Software and Technology}, UIST '19, p. 1093–1107. Association for Computing Machinery, New York, NY, USA, 2019. doi: {{%
10\hspace{.1pt}\discretionary{.}{%
}{.}\hspace{.4pt}1145\discretionary{/}{%
}{/}3332165\hspace{.1pt}\discretionary{.}{%
}{.}\hspace{.4pt}3347875}}


\bibitem{10.1111/j.1468-2958.2007.00299.x}
N.~Yee and J.~Bailenson.
\newblock {The Proteus Effect: The Effect of Transformed Self-Representation on Behavior}.
\newblock {\em Human Communication Research}, 33(3):271--290, 07 2007. doi: {{%
10\hspace{.1pt}\discretionary{.}{%
}{.}\hspace{.4pt}1111\discretionary{/}{%
}{/}j\hspace{.1pt}\discretionary{.}{%
}{.}\hspace{.4pt}1468\discretionary{%
}{-}{-}2958\hspace{.1pt}\discretionary{.}{%
}{.}\hspace{.4pt}2007\hspace{.1pt}\discretionary{.}{%
}{.}\hspace{.4pt}00299\hspace{.1pt}\discretionary{.}{%
}{.}\hspace{.4pt}x}}


\bibitem{web2023YT2}
{YouTube: @TheVirtualRealityShow}.
\newblock {Best Places to see in VRCHAT!}, 2023.

\bibitem{Yu2023}
K.~Yu, S.~Wen, W.~Xu, M.~Caon, N.~Baghaei, and H.-N. Liang.
\newblock {Cheer for me: effect of non-player character audience feedback on older adult users of virtual reality exergames}.
\newblock {\em Virtual Reality}, 27(3):1887--1903, 2023. doi: {{%
10\hspace{.1pt}\discretionary{.}{%
}{.}\hspace{.4pt}1007\discretionary{/}{%
}{/}s10055\discretionary{%
}{-}{-}023\discretionary{%
}{-}{-}00780\discretionary{%
}{-}{-}5}}


\bibitem{zhu2023}
J.~Zhu, R.~Kumaran, C.~Xu, and T.~Höllerer.
\newblock Free-form conversation with human and symbolic avatars in mixed reality.
\newblock In {\em 2023 IEEE International Symposium on Mixed and Augmented Reality (ISMAR)}, 2023.

\end{thebibliography}
\end{document}